\documentclass[10pt]{article}

\usepackage{amssymb}
\usepackage{graphicx}          
\usepackage{epsf}
\usepackage{graphics}
\usepackage{bm}       
\usepackage{epsfig}
\usepackage{amsmath}
\usepackage{soul,color}
\usepackage{epsfig}
\usepackage{apacite}
\usepackage{natbib}
\usepackage{hyperref}
\usepackage{booktabs}
\usepackage{longtable}
\usepackage{array}

\newcolumntype{M}[1]{>{\centering\arraybackslash}m{#1}}
\newcolumntype{N}{@{}m{0pt}@{}}

\usepackage{float}
\usepackage[nolists]{endfloat}

\usepackage{tikz} 
\usetikzlibrary{calc,shapes} 
\usetikzlibrary{arrows,snakes,backgrounds}
\usetikzlibrary{shapes,snakes}
\tikzstyle{arr}=[-latex,black]
\tikzstyle{latent}=[circle,draw,inner sep=0pt,minimum size=6mm]
\tikzstyle{manifest}=[rectangle,draw,inner sep=0pt,minimum size=6mm]
\tikzset{>=latex}

\def\baselinestretch{1.1}

\begin{document}
\title{Conditioning on the pre-test versus gain score modeling: revisiting the controversy in a multilevel setting}
\author{Bruno Arpino, Silvia Bacci, Leonardo Grilli,  Raffaele Guetto, Carla Rampichini}
\date{Dep. of Statistics, Computer science, Applications -
University of Florence (IT)}

\maketitle


\thispagestyle{empty}
\def\baselinestretch{1.2}

\begin{abstract}
We consider estimating the effect of a treatment on the progress of subjects tested both before and after treatment assignment. A vast literature compares the competing approaches of modeling the post-test score conditionally on the pre-test score versus modeling the difference, namely the gain score. Our contribution resides in analyzing the merits and drawbacks of the two approaches in a multilevel setting. This is relevant in many fields, for example education with students nested into schools. The multilevel structure raises peculiar issues related to the contextual effects and the distinction between individual-level and cluster-level treatment. We derive approximate analytical results and compare the two approaches by a simulation study. For an individual-level treatment our findings are in line with the literature, whereas for a cluster-level treatment we point out the key role of the cluster mean of the pre-test score, which favors the conditioning approach in settings with large clusters.  
\end{abstract}

\newpage

\setcounter{page}{1}

\section{Introduction}
\label{sec:intro}

We consider the problem to assess a treatment effect on an outcome repeatedly measured before and after the treatment assignment in observational studies. In particular, we focus on the educational setting, where the performance of a student is assessed by two achievement tests at different educational stages, and the target is to estimate the effect of a school-level or student-level treatment applied between the two tests, hence named pre-test and post-test. 

The evaluation of the treatment effect can be achieved by means of two approaches  \citep[see, for instance, ][]{bereiter:1963, lord:1963}. The first approach consists in estimating the effect of the treatment on the post-test score, conditionally on the pre-test score (\emph{conditioning} approach). In the  second approach, the analysis is carried out on the gain score (also known as change score), obtained as the difference between the post-test score and the pre-test score ({\em gain score} approach).

Disappointingly, the conditioning and gain score approaches may give contradictory results in observational studies and a treatment effect may be statistically significant or not depending on the chosen approach \citep[Lord's paradox; ][]{lord:1967}. The debate on which of the two approaches has to be preferred is still ongoing. Recently, \cite{kim:steiner:2016} and \cite{kim:steiner:2019} reconsidered this issue using graphical models to show the conditions under which an approach has to be preferred. 
Despite the important contribution of this and several other studies that we review later, the literature has overlooked the fact that test scores are typically collected in multilevel settings: standardized scores on students' performance represent the prototypical case with students nested within schools.
Indeed, in education, value added analysis is routinely carried out by means of multilevel models \citep{gold:2010}, mostly adopting the conditioning approach \citep{Raudenbush:2005}. 

The educational literature has long been interested in assessing how students' performance is affected by the characteristics and choices of students \citep{Webbink:2005, Nordenbo:2010} and the features of schools and teachers \citep{Deangelis:2019}. For instance, \cite{Heller:2018} estimated the effect of attending a private school instead of a public one on students' performance in mathematics and reading, showing that students in private schools have better results on both tests. As an example of an individual-level treatment, \cite{Hong:Raudenbush:2006} found a negative effect of being retained in kindergarten rather than being promoted on pupils' learning.

Value added analysis mainly aims at estimating overall school effects adjusted for observed characteristics of students and schools \citep{Timmermans:2015}. In our contribution, however, we do not focus on overall school effects, rather we intend to assess the effect of a given treatment, such as attending a public or private school. 
To this aim, we will compare the competing approaches of modeling the post-test conditional on the pre-test versus modeling the gain score, explicitly accounting for the multilevel structure of the data. The multilevel setting generates new scenarios. First of all, the treatment of interest may be at the individual level or at the cluster level. Moreover, the multilevel setting raises the issue of contextual effects, namely the cluster mean of a variable may have a non-null coefficient summarizing the effect of the context associated with the clustering. 

In addition, we will extend the results of \cite{kim:steiner:2019} in two directions not related to the multilevel setting: (\emph{i}) derivation of the bias formulas in the case of a binary treatment, e.g. private versus public school, and (\emph{ii}) further developments in the situation with a common measurement error affecting both pre-test and post-test scores.

The remaining of the paper is organized as follows. In Section \ref{sec:liter} we briefly summarize the debate on the use of conditioning versus gain score approaches. In Section (\ref{sec:method}) we discuss the two approaches in the single-level setting, with some extensions of the results of \cite{kim:steiner:2019}. In Section \ref{sec:multilevel} we compare the two approaches in a multilevel setting, whereas in Section \ref{sec:simulation} we devise a simulation study to assess the performance of the estimators. In Section \ref{sec:conclusion} we summarize our findings and give some directions for future research.

\section{Literature overview}
\label{sec:liter}
Traditionally, the debate about the performance of the conditioning and gain score approaches has favored the former, mainly because of the sensitivity of the gain score approach to the regression toward the mean and the consequent inexactness  
of gain scores in comparison with pre-test and post-test scores.

In the 90s, \cite{allison:1990} resumed the debate on the two methods partly disproving those widespread beliefs, and showing that the gain score method has to be considered superior when the treatment is subsequent to the pre-test and is uncorrelated with the time-varying components of the pre-test.

Afterwards, \cite{maris:1998} and, more recently, \cite{breukelen:2013} achieved some interesting results about the relation between the performance of the two approaches and the treatment assignment mechanism.  They   showed that  the conditioning approach has to be preferred to the gain score one when the treatment assignment is based on a randomized procedure or on the pre-test score. Under randomization, both approaches are unbiased but the conditioning one is more powerful, whereas under  assignment depending on the pre-test score only the conditioning approach is unbiased. These conclusions hold regardless of whether the pre-test suffers from measurement error \citep{maris:1998}. 

Although these  results seem to sustain the  conditioning approach, the consequences related with treatment assignment procedures based on pre-existing groups are anything but definitive, as discussed by \cite{breukelen:2013} that provided  a novel interpretation of the two approaches in terms of models for repeated measures. On the one hand, the conditioning approach yields unbiased estimates provided that pre-test group differences are absent, implying   the regression of the post-test scores toward a common mean if neither group is treated. On the other hand, the estimates of the gain score approach are unbiased if the pre-existing groups show the same changes when neither group is treated. Unfortunately, both assumptions are untestable. Additionally, no clear indication can be derived if the treatment is self-assigned on the basis of  individuals' preferences.

\cite{kim:steiner:2019} further contributed to the debate focusing on a causal inference perspective and considering the situation in which the treatment assignment is based on an unobservable variable (latent ability), as it often happens in observational studies.

Indeed, the causal inference literature makes it clear that the conditioning approach yields valid inferences under the  unconfoundedness assumption, namely  when there is conditional independence between treatment and potential outcomes \citep{Imbens:2015, arpino:aassve:2013}. On the other hand, the gain score approach entails taking the first difference of the outcome, thus it removes confounding if the confounders are time invariant and their effect is constant over time, the so called \emph{common trend assumption} \citep[e.g., ][]{lechner:2011}. 

Exploiting graphical models, \cite{kim:steiner:2019} derived formulas for the bias of causal effects estimators under the two approaches. In particular, they considered a  linear data generating model with constant effects across all units, where a  latent ability affects three observed variables, namely the treatment variable, the pre-test score and the post-test score. Additionally, the treatment variable affects the post-test score. In such a setting, a low pre-test reliability (measurement error) favors the gain score approach. However, the gain score approach relies on the common trend assumption, which cannot be tested with only two measures available on the test score. 
Violations of the common trend assumption would favor the conditioning approach. \cite{kim:steiner:2019} also considered more complex scenarios, in particular a direct effect of the pre-test score on the treatment variable, which complicates the assessment of the bias under the gain score approach and, consequently, the choice between the two approaches. 

The literature on the conditioning versus gain score approaches has mainly focused on unstructured (single-level) data, overlooking the fact that in the educational context the evaluation of performances after an intervention may be more complex due the hierarchical structure of data, that is students nested into  clusters such as classes and schools. In this setting, the treatment may intervene at the individual level (i.e., the treatment unit is the student) or at the cluster level (i.e., the treatment unit is the school or the class) and individual-level variables, such as the latent ability, may have a relevant contextual effect (i.e., an effect due to the aggregation of individuals in clusters). 

All these aspects  make the comparison between the conditioning and the gain score approaches in the multilevel setting more complex. 
In their recent paper \cite{kohler:2020} considered a multilevel setting, however they focus only on the cluster (school/class) level, ignoring the individual (student) level. In our knowledge, our paper is the first one investigating the treatment effect in a fully multilevel setting, simultaneously considering cluster and individual levels.

\section{Conditioning and gain score approaches in a single-level setting}
\label{sec:method}
In this section,  we first specify the model and briefly review the findings of \cite{kim:steiner:2019} (Section \ref{sec:basic}), obtained for a linear model with a continuous treatment. Then, we consider the bias for a binary treatment (Section \ref{sec:binary_treat}) and we make further considerations on the situation where a common measurement error affects both the pre-test and post-test scores (Section \ref{sec:common_error}).

\subsection{Basic models} \label{sec:basic}
We deal with a data structure where individuals ($i = 1, \ldots, n_j$) are nested into clusters ($j = 1, \ldots, J$).
In this section we describe the conditioning and the gain score approaches in a single-level observational setting, thus ignoring the clustering.  Notwithstanding, we denote the individuals with the double index $ij$ to avoid changing notation when extending the analysis to the multilevel setting (Section \ref{sec:multilevel}).

Let $A_{ij}$ be a latent variable summarizing the unobservable ability of individual (student) $i$ nested in cluster (school) $j$.  Moreover, let $Y_{1ij}$ and $Y_{2ij}$ be continuous variables for the observed scores on the pre-test and the post-test, respectively, for individual $i$ in cluster $j$; in turn,  the difference $G_{ij} = Y_{2ij} - Y_{1ij}$ is the gain score. We assume that $A_{ij}$ is an unobservable confounder that affects both $Y_{1ij}$  and $Y_{2ij}$. 
In addition, we denote by $Z_{ij}$ a treatment at the individual level assigned after the pre-test. In \cite{kim:steiner:2019}, $Z_{ij}$  is a continuous variable, whereas we also consider the case where $Z_{ij}$ is a binary variable equal to 1 if individual $i$ is treated and 0 otherwise (e.g., the student enrolls in a public vs. a private school). 
Finally, we introduce the random terms $e_{ij}$ and $v_{ij}$ to account for the errors in the measurement of the individual ability through the pre-test and post-test, respectively.

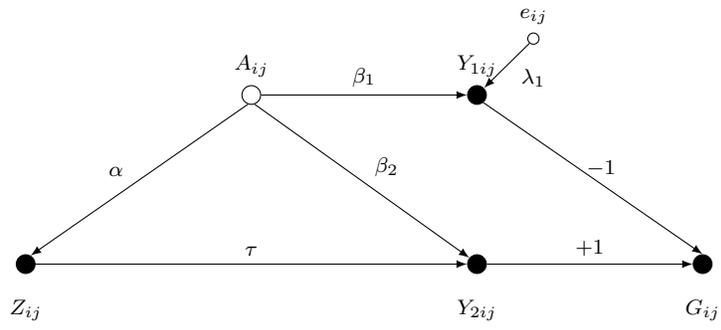
\begin{figure}[!ht]
\centering
\small
\begin{tikzpicture}[scale=1.5]
\node  (A) at (-2.0,0.0) [shape=circle, draw, inner sep=2.5pt, label = {\footnotesize{$\;\;A_{ij}\;\;$}}] {};
\node (Z) at (-4.0,-1.5) [shape=circle, draw, fill, inner sep=2.5pt, label = {[label distance = -1.0cm]\footnotesize{$\;\;Z_{ij}\;\;$}}] {};
\node (P) at (0.0,0.0)  [shape=circle, draw, fill, inner sep=2.5pt, label = {\footnotesize{$\;\;Y_{1ij}\;\;$}}] {};
\node (Y) at (0.0,-1.5)  [shape=circle, draw, fill, inner sep=2.5pt, label = {[label distance = -1.0cm]\footnotesize{$\;\;Y_{2ij}\;\;$}}] {};
\node (G) at (2.0,-1.5) [shape=circle, draw, fill,  inner sep=2.5pt, label = {[label distance = -1.0cm]\footnotesize{$\;\;G_{ij}\;\;$}}] {};
\node (E) at (0.5,0.5)  [shape=circle, draw, inner sep=1.5pt, label = {\footnotesize{$\;\;e_{ij}\;\;$}}] {};
\draw [arr, ->]  (A.east) -- (P.west);
\draw [arr, ->]  (A.290) -- (Y.140);
\draw [arr, ->]  (A.250) -- (Z.60);
\draw [arr, ->]  (Z.east) -- (Y.west);
\draw [arr, ->]  (Y.east) -- (G.west);
\draw [arr, ->]  (P.310) -- (G.north);
\draw [arr, ->]  (E.230) -- (P.45);
\node [above] at (-3.2, -0.8) {\footnotesize{$\alpha$}};
\node [above] at (-2.0, -1.5) {\footnotesize{$\tau$}};
\node [above] at (-1.0, 0.0) {\footnotesize{$\beta_1$}};
\node [above] at (1.0, -1.5) {\footnotesize{$+1$}};
\node [above] at (-0.8, -0.8) {\footnotesize{$\beta_2$}};
\node [above] at (1.1, -0.8) {\footnotesize{$-1$}};
\node [above] at (0.5, 0.0) {\footnotesize{$\lambda_1$}}; 
\end{tikzpicture}
\caption{\em Path diagram for the data generating model in a single-level setting with measurement error on the pre-test (full circle: observed variable; empty circle: unobserved variable).}
\label{graph:1}
\end{figure}

Figure \ref{graph:1} reports the path diagram for the basic model in the single-level framework. This is the same as in \cite{kim:steiner:2019} with the only difference of the additional subscript $j$. 
In Figure \ref{graph:1}, $\beta_1$ and $\beta_2$ denote the effect of the latent ability on the pre-test and the post-test, respectively, whereas $\tau$ represents the effect of the treatment on the post-test, which is the target of our analysis. Moreover, $\alpha$ accounts for the influence of the ability on the treatment and $\lambda_1$ accounts for the measurement error on the pre-test.    

Consistently with the path diagram of Figure \ref{graph:1}, the pre-test score is assumed to be generated by a linear model:
\begin{equation}
Y_{1ij} = \mu_1 + \beta_1 A_{ij} + \lambda_1 e_{ij},
\label{eq:gen0:P}
\end{equation}
where $\mu_1$ is the intercept and $\lambda_1 e_{ij}$ is the measurement error, regulated by the parameter $\lambda_1 \ge 0$. If $\lambda_1 = 0$ there is no measurement error, thus $Y_{1ij}$ is a perfect measure of the latent ability up to a scale factor. Without loss of generality, we assume that both the individual ability and the error term  have zero mean and unit variance, that is, $E(A_{ij}) = E(e_{ij}) = 0$  and  $Var(A_{ij}) = Var(e_{ij}) = 1$. Then, $Var(Y_{1ij}) = \beta_1^2 + \lambda_1^2$.

In addition, also the post-test score is assumed to be generated by a linear model:
\begin{equation} 
Y_{2ij} = \mu_2 + \beta_2 A_{ij} +  \tau Z_{ij}  + v_{ij}, 
\label{eq:gen0:Y}
\end{equation}
with $\mu_2$ being a constant term and $v_{ij}$ is an error component with zero mean and constant variance.

The focus of our analysis is the estimation of the treatment effect on the post-test score represented by the coefficient $\tau$ in model \eqref{eq:gen0:Y}. Because model \eqref{eq:gen0:Y} involves the ability $A_{ij}$, which is not observed, it cannot be directly used to estimate $\tau$. As discussed in Section \ref{sec:liter}, in the literature there exist two alternative approaches to deal with this issue: the conditioning approach and the gain score approach.

The {\em conditioning} approach, which is the standard way to remove confounding, entails using the pre-test score $Y_{1ij}$ in place of the latent ability $A_{ij}$ in equation \eqref{eq:gen0:Y}, thus fitting the following linear model:
\begin{equation}
Y_{2ij} = \mu^*  + \beta_2^{*} Y_{1ij}  +  \tau^* Z_{ij} + v_{ij}^*.
\label{eq:est_cond:Y}
\end{equation}
The drawback of this approach is the attenuation of the regression coefficient of the pre-test score $Y_{1ij}$ due to measurement error. Therefore, conditioning on $Y_{1ij}$ only partially adjusts for the latent ability, causing a bias in the estimation of the treatment effect $\tau$ \citep{breukelen:2013, kim:steiner:2019}. The bias depends on the magnitude of the measurement error, regulated by the parameter $\lambda_1$, and vanishes only in absence of measurement error ($\lambda_1 = 0$).

The expectation of the OLS estimator of the treatment effect $\tau$ in the presence of measurement error on $Y_{1ij}$ is derived by \cite{kim:steiner:2019}: 

\begin{align} 
b_{Y_2Z.Y_1} 
& = \tau  + \frac{\alpha \beta_2 \lambda_1^2  }{Var(Z_{ij})(\beta_1^2+\lambda_1^2)-\alpha^2\beta_1^2}
\label{eq:biasPnew}
\\
& = \tau + \frac{\alpha \beta_2(1-\rho)}{Var(Z_{ij})-\alpha^2\rho}, \label{eq:biasP}
\end{align}
where $\rho$ is the reliability of the pre-test
\begin{equation}
\rho = \frac{\beta_1^2Var(A_{ij})}{Var(Y_{1ij})}  = \frac{\beta_1^2}{\beta_1^2 + \lambda_1^2}.
\label{eq:rho}
\end{equation}
Formula (\ref{eq:biasP}) shows that the bias decreases as the reliability $\rho$ increases, vanishing when $\rho=1$ (which is achieved for $\lambda_1 = 0$, i.e., no measurement error). It is worth to note that formula \eqref{eq:biasP} is derived assuming a continuous treatment $Z_{ij}$. This is not very common in practice, thus in Section \ref{sec:binary_treat} we extend the formula to the case of a binary treatment.

On the other hand, the {\em gain score} approach consists in assessing the effect of the treatment on the gain score, namely the difference between the post-test score (equation \ref{eq:gen0:Y}) and the pre-test score (equation \ref{eq:gen0:P}):
\begin{align}\nonumber
G_{ij} & = Y_{2ij} - Y_{1ij} \\\nonumber
 & = \left(\mu_2 + \beta_2 A_{ij} +  \tau Z_{ij} + v_{ij}\right) - \left(\mu_1 + \beta_1 A_{ij} + \lambda_1 e_{ij}\right) \\
 & = (\mu_2 - \mu_1) + (\beta_2 - \beta_1) A_{ij} + \tau Z_{ij} +  (v_{ij}- \lambda_1 e_{ij}).
 \label{eq:gain}
 \end{align}
 
If the latent ability has the same effect on both the pre-test and the post-test (common trend assumption), that is
$$
\beta_2  = \beta_1,
$$  
then $A_{ij}$ in equation \eqref{eq:gain} vanishes and the gain score model becomes
$$
G_{ij} = (\mu_2 - \mu_1) +  \tau Z_{ij} +  (v_{ij}- \lambda_1 e_{ij}),
$$
thus the treatment effect $\tau$ is estimable without bias, regardless of the measurement error in the pre-test.
As  noted by \cite{kim:steiner:2019}, the pre-test score $Y_{1ij}$ should not be inserted as a covariate in the gain score model, because this would bring the measurement error back.

In summary, in the setting illustrated in Figure \ref{graph:1}, the conditioning approach is advantageous when the pre-test is measured without error; on the opposite, the gain score approach should be preferred when the common trend is a realistic assumption, regardless of the reliability of the pre-test.  

\subsection{Binary treatment}
\label{sec:binary_treat}
The bias due to measurement error on $Y_{1ij}$ in equation (\ref{eq:biasP}) is derived by \cite{kim:steiner:2019} in the case of a continuous treatment $Z_{ij}$. However, in most cases the treatment $Z_{ij}$ is binary, so the bias in formula (\ref{eq:biasP}) should be changed accordingly.  \cite{kim:steiner:2016} in Appendix B provided this formula in the special case of a latent linear model with Normal errors, while in the following we consider a general nonlinear case.

Regardless of the nature of $Z_{ij}$, the OLS estimator of $Y_{2ij}$ on $Z_{ij}$ given $Y_{1ij}$ has expectation
\begin{equation} 
\label{eq:biasZP}
b_{Y_2Z.Y_1}  = \tau +\frac{\beta_2(1-\rho)E(A_{ij}  Z_{ij})}{Var(Z_{ij})-\rho E(A_{ij}  Z_{ij})^2}.
\end{equation}
Recalling that $E(A_{ij})=0$ and $Var(A_{ij})=1$ and assuming that $Z_{ij}$ is a continuous treatment generated by a linear model $Z_{ij} = \alpha A_{ij}+\varepsilon_Z$, it follows that $E(A_{ij} Z_{ij}) = Cov(A_{ij}, Z_{ij}) = \alpha$, thus formula (\ref{eq:biasZP}) reduces to formula (\ref{eq:biasP}).

Now let us consider a binary treatment $Z_{ij}$ depending on the latent ability $A_{ij}$ through an arbitrary nonlinear relationship. We define $\pi_{ij} = Pr(Z_{ij} = 1 | A_{ij}) = E(Z_{ij} | A_{ij})$ and $\pi = E(\pi_{ij})$, namely $\pi$ is the expectation of $\pi_{ij}$ over the distribution of $A_{ij}$.
For example, the relationship between $Z_{ij}$ and $A_{ij}$ might be logistic:

\begin{equation}
\label{eq:logit}
\pi_{ij} = Pr(Z_{ij} = 1 \mid A_{ij})=\frac{1}{1+\exp(-(\delta+\alpha A_{ij}))}.
\end{equation}

Given that $Var(Z_{ij}) = \pi (1-\pi)$ and
\begin{align*}
E(A_{ij}  Z_{ij}) & = E(A_{ij}  Z_{ij} \mid Z_{ij} = 0)(1-\pi) + \\
    & + E(A_{ij}  Z_{ij} \mid Z_{ij} = 1)\pi \\
    & = E(A_{ij} \mid Z_{ij} = 1)\pi,
\end{align*}
the bias formula \textbf{(\ref{eq:biasZP})} in the case of a binary treatment becomes
\begin{equation} 
\label{eq:biasZP2}
b_{Y_2Z.Y_1}  = \tau +\frac{\beta_2(1-\rho)E(A_{ij} \mid Z_{ij} = 1)}{(1 -\pi)-\rho \pi [E(A_{ij} \mid Z_{ij} = 1)]^2}.
\end{equation}

Note that in the binary case the association between treatment and ability is summarized by the conditional mean of the ability across treatment levels, instead of the regression parameter $\alpha$. Nonetheless, the bias still depends on the reliability and it vanishes for a unit reliability. 

\subsection{Common measurement error}
\label{sec:common_error}
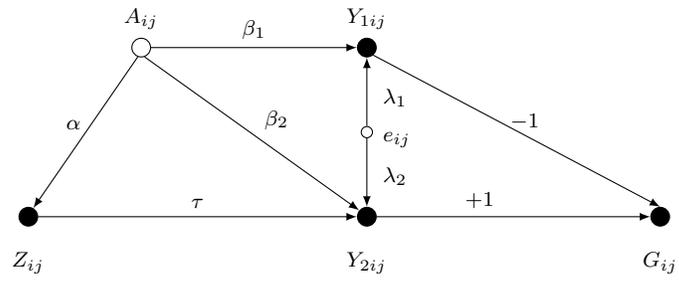
\begin{figure}[!ht]
\centering
\small
\begin{tikzpicture}[scale=1.5]
\node  (A) at (-2.0,0.0) [shape=circle, draw, inner sep=2.5pt, label = {\footnotesize{$\;\;A_{ij}\;\;$}}] {};
\node (Z) at (-3.0,-1.5) [shape=circle, draw, fill, inner sep=2.5pt, label = {[label distance = -1.0cm]\footnotesize{$\;\;Z_{ij}\;\;$}}] {};
\node (P) at (0.0,0.0)  [shape=circle, draw, fill, inner sep=2.5pt, label = {\footnotesize{$\;\;Y_{1ij}\;\;$}}] {};
\node (Y) at (0.0,-1.5)  [shape=circle, draw, fill, inner sep=2.5pt, label = {[label distance = -1.0cm]\footnotesize{$\;\;Y_{2ij}\;\;$}}] {};
\node (G) at (2.6,-1.5)  [shape=circle, draw, fill,  inner sep=2.5pt, label = {[label distance = -1.0cm]\footnotesize{$\;\;G_{ij}\;\;$}}] {};
\node (E) at (0.0,-0.75) [shape=circle, draw, inner sep=1.5pt, label = {[xshift=0.4cm, yshift=-0.4cm]\footnotesize{$\;\;e_{ij}\;\;$}}] {};
\draw [arr, ->]  (A.east) -- (P.west);
\draw [arr, ->]  (A.290) -- (Y.140);
\draw [arr, ->]  (A.250) -- (Z.60);
\draw [arr, ->]  (Z.east) -- (Y.west);
\draw [arr, ->]  (Y.east) -- (G.west);
\draw [arr, ->]  (P.310) -- (G.north);
\draw [arr, ->]  (E.north) -- (P.south);
\draw [arr, ->]  (E.south) -- (Y.north);
\node [above] at (-2.6, -0.8) {\footnotesize{$\alpha$}};
\node [above] at (-1.5, -1.5) {\footnotesize{$\tau$}};
\node [above] at (-1.0, 0.0)  {\footnotesize{$\beta_1$}};
\node [above] at (1.0, -1.5)  {\footnotesize{$+1$}};
\node [above] at (-0.8, -0.8) {\footnotesize{$\beta_2$}};
\node [above] at (1.4, -0.8)  {\footnotesize{$-1$}};
\node [above] at (0.25, -0.6) {\footnotesize{$\lambda_1$}}; 
\node [above] at (0.25, -1.3) {\footnotesize{$\lambda_2$}}; 
\end{tikzpicture}
\caption{\em Path diagram for the data generating model in a single-level setting with measurement error common to pre-test and post-test (full circle: observed variable; empty circle: unobserved variable).}
\label{graph:1b}
\end{figure}

Often the pre-test and post-test are administered through the same instrument (e.g., a standardized test). In such a case, the pre-test and post-test scores are likely to be affected by a common measurement error. This situation is represented in Figure \ref{graph:1b}, which generalizes Figure \ref{graph:1} with the introduction of the unobserved variable $e_{ij}$ affecting both the pre-test and post-test scores with parameters $\lambda_1$ and $\lambda_2$, respectively.
According to the path diagram of Figure \ref{graph:1b}, the model of the post-test score becomes
\begin{equation} 
Y_{2ij} = \mu_2 + \beta_2 A_{ij} +  \tau Z_{ij}  + \lambda_2 e_{ij}+v_{ij}, 
\label{eq:gen0:Ycollider}
\end{equation}
which reduces to equation \eqref{eq:gen0:Y} when  $\lambda_2 = 0$. The model of the pre-test score \eqref{eq:gen0:P} is unchanged.  
Due to the common factor $e_{ij}$, the error terms of the two test scores are correlated, specifically $Cov(\lambda_1  e_{ij}, \lambda_2  e_{ij}+v_{ij})=\lambda_1 \lambda_2  Var(e_{ij})$.

\cite{kim:steiner:2019} noted that, although the common measurement error $e_{ij}$ generates a correlated error structure, it does not affect the bias in gain score approach, while it yields an additional bias term in the conditioning approach. In detail, the bias formula \eqref{eq:biasZP} generalizes as follows:
\begin{align} \nonumber
\label{eq:bias0}
b_{Y_2Z.Y_1} & = \tau +\frac{\beta_2 (1-\rho)E(A_{ij} Z_{ij})}{Var(Z_{ij}) - \rho E(A_{ij}Z_{ij})^2} \\\nonumber
& - \frac{\beta_1\lambda_1\lambda_2E(A_{ij} Z_{ij})}{\left[Var(Z_{ij}) - \rho E(A_{ij}Z_{ij})^2\right] Var(Y_{1ij})} \\
&  = \tau +\frac{E(A_{ij} Z_{ij})\left[\beta_2 (1-\rho)Var(Y_{1ij}) - \beta_1\lambda_1\lambda_2\right]}{\left[Var(Z_{ij}) - \rho E(A_{ij}Z_{ij})^2\right] Var(Y_{1ij})}.
\end{align}

Given the expression of $\rho$ in equation \eqref{eq:rho}, formula \eqref{eq:bias0} simplifies to
\begin{equation} 
\label{eq:collbias2}
b_{Y_2Z.Y_1}  = \tau  + \frac{\alpha (\beta_2 \lambda_1^2 -  \beta_1\lambda_1\lambda_2)}{Var(Z_{ij})(\beta_1^2+\lambda_1^2)-\alpha^2\beta_1^2},
\end{equation}
when $Z_{ij}$ is a continuous treatment \citep{kim:steiner:2019}, whereas it becomes
\begin{equation} 
\label{eq:biasZP3}
b_{Y_2Z.Y_1}  =   \tau +\frac{E(A_{ij} \mid Z_{ij} = 1)(\beta_2\lambda_1^2 -\beta_1\lambda_1\lambda_2)}{(1 -\pi)(\beta_1^2+\lambda_1^2)- \pi \left[E(A_{ij} \mid Z_{ij} = 1)\right]^2 \beta_1^2},
\end{equation}
when $Z_{ij}$ is a binary treatment.

We outline that, regardless of the nature of $Z_{ij}$, in case of common measurement error the bias in the estimation of $\tau$ has $\beta_2\lambda_1^2 -\beta_1\lambda_1\lambda_2$ in the numerator. Therefore, under the common trend assumption ($\beta_1 = \beta_2$), if the effect of the common measurement error $e_{ij}$ is the same for the pre-test and post-test ($\lambda_1=\lambda_2$), the bias component in equations \eqref{eq:collbias2} and \eqref{eq:biasZP3} vanishes. In general, a common measurement error is beneficial for the conditioning approach since it reduces the bias, as it can be seen by comparing equations (\ref{eq:biasPnew}) and (\ref{eq:collbias2}), given that the $\beta$s and the $\lambda$s are typically positive.

\section{Conditioning and gain score approaches in a multilevel setting}
\label{sec:multilevel}
When the data have a multilevel structure, the framework outlined in the previous section has to be extended to handle new issues.

First of all, it is essential to distinguish the case of an individual-level treatment $Z_{ij}$ (e.g., the treatment unit is the student) from the case of a cluster-level treatment $Z_{j}$ (e.g., the treatment unit is the school). In the exposition we consider the two cases separately since they have different implications for the analysis.

A common issue in multilevel settings is the presence of contextual effects of individual-level variables. Specifically, it may be that the effects of the ability on the pre-test and post-test scores are different at the individual and cluster levels, so that the scores may depend also on the cluster mean of the ability.

A further complication concerns the genesis of the clusters. We focus on the case where the clusters at the time of the post-test are formed before treatment assignment and they are the same at pre-test and post-test. In this situation, the cluster mean of the ability can affect both the treatment and the pre-test score, in addition to the post-test score. For example, this is the case of the analysis of \cite{Hong:Raudenbush:2006} on the effect of retention to kindergarten on pupils' performance. 

In all scenarios, the unobserved ability is a possible confounder which is measured with error through the pre-test. Therefore, the pre-test score is the key to eliminate confounding, either by conditioning or by computing the gain score.

\subsection{Individual-level treatment}
\label{sec:individual-treat}
If the treatment acts at the individual level, for instance the students are individually assigned to an experimental didactic program, the path diagram of Figure \ref{graph:1b} generalizes as shown in Figure \ref{graph:2a}, where the cluster mean of the latent ability $\bar{A}_j = \frac{1}{n_j}\sum_{i = 1}^{n_j}A_{ij}$ is a new variable affecting both the pre-test score $Y_{1ij}$ and the post-test score $Y_{2ij}$ (contextual effects).
\begin{figure}[!ht]
\centering
\small
\begin{tikzpicture}[scale=1.5]
\node (A) at (-2.0,0.0) [shape=circle, draw, inner sep=2.5pt, label = {\footnotesize{$\;\;A_{ij}\;\;$}}] {};
\node (Z) at (-3.0,-1.5) [shape=circle, draw, fill, inner sep=2.5pt, label = {[label distance = -1.0cm]\footnotesize{$\;\;Z_{ij}\;\;$}}] {};
\node (P) at (0.0,0.0)  [shape=circle, draw, fill, inner sep=2.5pt, label = {\footnotesize{$\;\;Y_{1ij}\;\;$}}] {};
\node (Y) at (0.0,-1.5)  [shape=circle, draw, fill, inner sep=2.5pt, label = {[label distance = -1.0cm]\footnotesize{$\;\;Y_{2ij}\;\;$}}] {};
\node (G) at (2.6,-1.5) [shape=circle, draw, fill,  inner sep=2.5pt, label = {[label distance = -1.0cm]\footnotesize{$\;\;G_{ij}\;\;$}}] {};
\node (E) at (0.0,-0.75)  [shape=circle, draw, inner sep=1.5pt, label = {[xshift=0.4cm, yshift=-0.4cm]\footnotesize{$\;\;e_{ij}\;\;$}}] {};
\node  (Abar) at (-1.0,0.8) [shape=circle, draw, inner sep=2.5pt, label = {\footnotesize{$\;\;\bar{A}_{j}\;\;$}}] {};
\draw [arr, ->]  (A.east) -- (P.west);
\draw [arr, ->]  (A.290) -- (Y.140);
\draw [arr, ->]  (A.250) -- (Z.60);
\draw [arr, ->]  (Z.east) -- (Y.west);
\draw [arr, ->]  (Y.east) -- (G.west);
\draw [arr, ->]  (P.310) -- (G.north);
\draw [arr, ->]  (E.north) -- (P.south);
\draw [arr, ->]  (E.south) -- (Y.north);
\draw [arr, ->]  (A.45) -- (Abar.230);
\draw [arr, ->]  (Abar.290) -- (P.110);
\draw [arr, ->]  (Abar.290) -- (Y.100);
\node [above] at (-2.6, -0.8) {\footnotesize{$\alpha$}};
\node [above] at (-1.5, -1.5) {\footnotesize{$\tau$}};
\node [above] at (-1.0, 0.0) {\footnotesize{$\beta_1$}};
\node [above] at (1.0, -1.5) {\footnotesize{$+1$}};
\node [above] at (-0.8, -0.8) {\footnotesize{$\beta_2$}};
\node [above] at (1.4, -0.8) {\footnotesize{$-1$}};
\node [above] at (0.25, -0.6) {\footnotesize{$\lambda_1$}}; 
\node [above] at (0.25, -1.3) {\footnotesize{$\lambda_2$}}; 
\node [above] at (-0.3, 0.4) {\footnotesize{$\psi_1$}};
\node [above] at (-0.2, -0.6) {\footnotesize{$\psi_2$}};
\node [above] at (-1.6, 0.4) {\footnotesize{$cm(\cdot)$}};
\end{tikzpicture}
\caption{\em Path diagram for the data generating model in a multilevel setting with an individual-level treatment. Clusters formed before treatment assignment; measurement error common to pre-test and post-test (full circle: observed variable; empty circle: unobserved variable; $cm(\cdot)$: cluster mean).}
\label{graph:2a} 
\end{figure}
We denote with $\beta_1$ the within effect and with $\psi_1$ the contextual effect of the ability on the pre-test score, whereas $\beta_2$ and $\psi_2$ are the corresponding effects on the post-test score. Consequently, the generating model for $Y_{1ij}$ of equation (\ref{eq:gen0:P}) becomes
\begin{equation} \label{eq:gen1:P} 
Y_{1ij}  =  \mu_1  + \beta_1 A_{ij}  + \psi_1  \bar{A}_{j} +   u_{1j}+\lambda_1 e_{ij},
\end{equation} 
while the generating model for $Y_{2ij}$ of equation (\ref{eq:gen0:Ycollider}) becomes
\begin{equation} \label{eq:gen1:Y} 
Y_{2ij}  =  \mu_{2} + \beta_2 A_{ij}   + \psi_2  \bar{A}_{j}+  \tau Z_{ij}  + u_{2j}+   \lambda_2 e_{ij} + v_{ij}.
\end{equation}
The random intercept models (\ref{eq:gen1:P}) and (\ref{eq:gen1:Y}) have cluster-specific intercepts 
composed of a fixed component, $\mu_1  + \psi_1  \bar{A}_{j}$ and $\mu_2 + \psi_2  \bar{A}_{j}$, and a random component, $u_{1j}$ and $u_{2j}$, respectively. The cluster-level errors $u_{1j}$ and $u_{2j}$ are independently and identically distributed according to a bivariate distribution with zero means and unconstrained covariance matrix. 
In most applications, we expect $u_{1j}$ and $u_{2j}$ to be highly correlated. 
 
Note that in models (\ref{eq:gen1:P}) and (\ref{eq:gen1:Y}) the two components of the ability, namely $A_{ij}$ and $\bar{A}_{j}$, are correlated. For certain derivations, it is useful to reparametrize the models in terms of uncorrelated components of the ability. 
Indeed, the ability can be decomposed as ${A}_{ij} = \tilde{A}_{ij} + \bar{A}_{j}$, where $\tilde{A}_{ij}=A_{ij} - \bar{A}_{j}$ is the  deviation of the individual ability from its cluster mean. By construction, the within ability $\tilde{A}_{ij}$ and the between ability $\bar{A}_{j}$ are uncorrelated. Therefore, models (\ref{eq:gen1:P}) and (\ref{eq:gen1:Y}) can be reparametrized as follows: 
\begin{eqnarray} 
Y_{1ij}  = \mu_{1} + \beta_1^W \tilde{A}_{ij}  +\beta_1^B  \bar{A}_{j}  +   u_{1j}+ \lambda_1 e_{ij}, \label{eq:gen1:P2}  \\
  Y_{2ij}  = \mu_{2} + \beta_2^W \tilde{A}_{ij}  +\beta_2^B  \bar{A}_{j}  + \tau Z_{ij}+   u_{2j}+  \lambda_2 e_{ij} + v_{ij}, \label{eq:gen1:Y2}
\end{eqnarray} 
where $\beta_1^W=\beta_1$ and $\beta_2^W=\beta_2$ are the within effects, whereas $\beta_1^B=\beta_1^W+\psi_1$ and $\beta_2^B=\beta_2^W+\psi_2$ and the between effects.

As the treatment is at individual level, the source of bias in the estimation of the treatment effect $\tau$ is the same as in the single-level setting, namely the unobserved confounding due to the latent ability $A_{ij}$. Therefore, the solutions represented by the conditioning and gain score approaches are still valid, though they have to be implemented through multilevel models in order to get correct inferences.

In the conditioning approach an estimable version of model \eqref{eq:gen1:Y} is obtained by replacing $A_{ij}$ with the pre-test score $Y_{1ij}$ and $\bar{A}_j$ with the cluster-mean pre-test score $\bar{Y_1}_j = \frac{1}{n_j}\sum_{i = 1}^{n_j}Y_{1ij}$. However, in the data generating model (Figure \ref{graph:2a}) the cluster-mean latent ability $\bar{A}_j$ is not a confounder, thus the cluster-mean pre-test score $\bar{Y_1}_j$ can be omitted from the model without consequences on the bias of treatment effect $\tau$. This is important, since if the pre-test is affected by measurement error ($\lambda_1 \ne 0$) then $\bar{Y_1}_j$ is an error-prone measure of $\bar{A}_j$ that would suffer from attenuation. To summarize, in the multilevel setting the conditioning approach is implemented by fitting a random intercept model where $Y_{2ij}$ is regressed on $Z_{ij}$ and $Y_{1ij}$.

For the gain score approach the estimation model is obtained as difference between equations \eqref{eq:gen1:Y} and \eqref{eq:gen1:P}: 
\begin{equation} \label{eq:est_gain1}
G_{ij}  = \mu + (\beta_2 - \beta_1) A_{ij}  +  (\psi_2 - \psi_1) \bar{A}_{j}+ \tau Z_{ij} +  (u_{2j} - u_{1j})+\varepsilon_{ij},
\end{equation}
where $G_{ij} = Y_{2ij} - Y_{1ij}$, $\mu=\mu_{2} - \mu_{1}$ and $\varepsilon_{ij}=\lambda_2 e_{ij} + v_{ij} - \lambda_1 e_{ij}$.

Therefore, the gain score approach is implemented by fitting a random intercept linear model where $G_{ij}$ is regressed on $Z_{ij}$.
The novelty of a multilevel setting is that the common trend assumption has two parts: the individual-level part $\beta_2 = \beta_1$ let $A_{ij}$ vanish; the cluster-level part $\psi_2 = \psi_1$ let $\bar{A}_{j}$ vanish.
For unbiased estimation of the treatment effect $\tau$, the two parts of the assumption have a different role depending on the level of the treatment. 
If the treatment acts at the individual level ($Z_{ij}$), then $\beta_2 = \beta_1$ is required, while $\psi_2 = \psi_1$ is not required since $\bar{A}_{j}$ is not a confounder. 
On the other hand, if the treatment acts at the cluster level ($Z_{j}$), then both $\beta_2 = \beta_1$ and $\psi_2 = \psi_1$ are required. This case is treated next.

\subsection{Cluster-level treatment}
\label{sec:cluster-treat}
If the treatment acts at the cluster level, for example all students of a subset of schools are assigned to an experimental didactic program, the path diagram of Figure \ref{graph:1b} generalizes as shown in Figure \ref{graph:2b}. As in the case of an individual-level treatment (Figure \ref{graph:2a}), the cluster mean of the latent ability $\bar{A}_{j}$ affects both the pre-test score $Y_{1ij}$ and the post-test score $Y_{2ij}$ (contextual effects). However, in Figure \ref{graph:2b} the cluster mean of the latent ability $\bar{A}_{j}$ also affect the treatment variable $Z_{j}$.

\begin{figure}[!ht]
\centering
\small
\begin{tikzpicture}[scale=1.5]
\node (A) at (-2.0,0.0) [shape=circle, draw, inner sep=2.5pt, label = {\footnotesize{$\;\;\bar{A}_{j}\;\;$}}] {};
\node (Z) at (-3.0,-1.5) [shape=circle, draw, fill, inner sep=2.5pt, label = {[label distance = -1.0cm]\footnotesize{$\;\;Z_{j}\;\;$}}] {};
\node (P) at (0.0,0.0)  [shape=circle, draw, fill, inner sep=2.5pt, label = {\footnotesize{$\;\;Y_{1ij}\;\;$}}] {};
\node (Y) at (0.0,-1.5)  [shape=circle, draw, fill, inner sep=2.5pt, label = {[label distance = -1.0cm]\footnotesize{$\;\;Y_{2ij}\;\;$}}] {};
\node (G) at (2.6,-1.5) [shape=circle, draw, fill,  inner sep=2.5pt, label = {[label distance = -1.0cm]\footnotesize{$\;\;G_{ij}\;\;$}}] {};
\node (E) at (0.0,-0.75)  [shape=circle, draw, inner sep=1.5pt, label = {[xshift=0.4cm, yshift=-0.4cm]\footnotesize{$\;\;e_{ij}\;\;$}}] {};
\node  (Abar) at (-1.0,0.8) [shape=circle, draw, inner sep=2.5pt, label = {\footnotesize{$\;\;A_{ij}\;\;$}}] {};
\draw [arr, ->]  (A.east) -- (P.west);
\draw [arr, ->]  (A.290) -- (Y.140);
\draw [arr, ->]  (A.250) -- (Z.60);
\draw [arr, ->]  (Z.east) -- (Y.west);
\draw [arr, ->]  (Y.east) -- (G.west);
\draw [arr, ->]  (P.310) -- (G.north);
\draw [arr, ->]  (E.north) -- (P.south);
\draw [arr, ->]  (E.south) -- (Y.north);
\draw [arr, ->]  (Abar.230) -- (A.45);
\draw [arr, ->]  (Abar.290) -- (P.110);
\draw [arr, ->]  (Abar.290) -- (Y.100);
\node [above] at (-2.6, -0.8) {\footnotesize{$\alpha$}};
\node [above] at (-1.5, -1.5) {\footnotesize{$\tau$}};
\node [above] at (-1.0, 0.0) {\footnotesize{$\psi_1$}};
\node [above] at (1.0, -1.5) {\footnotesize{$+1$}};
\node [above] at (-0.8, -0.8) {\footnotesize{$\psi_2$}};
\node [above] at (1.4, -0.8) {\footnotesize{$-1$}};
\node [above] at (0.25, -0.6) {\footnotesize{$\lambda_1$}}; 
\node [above] at (0.25, -1.3) {\footnotesize{$\lambda_2$}}; 
\node [above] at (-0.3, 0.4) {\footnotesize{$\beta_1$}};
\node [above] at (-0.2, -0.6) {\footnotesize{$\beta_2$}};
\node [above] at (-1.6, 0.4) {\footnotesize{$cm(\cdot)$}};
\end{tikzpicture}
\caption{\em Path diagram for the data generating model in a multilevel setting with a cluster-level treatment. Clusters formed before treatment assignment; measurement error common to pre-test and post-test (full circle: observed variable; empty circle: unobserved variable; $cm(\cdot)$: cluster mean).}
\label{graph:2b}
\end{figure}

In the case of a cluster-level treatment, the model equations of the test scores $Y_{1ij}$ and $Y_{2ij}$ are still expressed by equations \eqref{eq:gen1:P} and \eqref{eq:gen1:Y}, except that in equation \eqref{eq:gen1:Y} the treatment variable $Z_{ij}$ has to be replaced by  $Z_{j}$, namely
\begin{equation} \label{eq:gen1:Y2} 
Y_{2ij} =  \mu_{2} + \beta_2 A_{ij}   + \psi_2  \bar{A}_{j} +  \tau Z_{j}  + u_{2j} +  \lambda_2 e_{ij} + v_{ij}. \end{equation}

The path diagram in Figure \ref{graph:2b} shows that the cluster mean of the latent ability $\bar{A}_{j}$ acts as a confounder of the effect of $Z_{j}$, therefore the implementation of the conditioning approach requires to fit a random intercept model with $Y_{2ij}$ regressed on $\bar{Y}_{1j}$, in addition to $Z_{j}$ and $Y_{1ij}$. 

The gain score approach is implemented, as usual, by computing the difference between the post-test and pre-test scores. In this case, subtracting equation \eqref{eq:gen1:P} from equation \eqref{eq:gen1:Y2} yields:

\begin{equation} \label{eq:est_gain2}
G_{ij}  =  \mu + (\beta_2 - \beta_1) A_{ij} + (\psi_2 - \psi_1) \bar{A}_{j} + \tau Z_{j}  +  (u_{2j} - u_{1j}) +  \varepsilon_{ij}.
\end{equation}
As a consequence of the confounding effect of $\bar{A}_{j}$ due to the treatment acting at cluster level, for unbiased estimation of the treatment effect $\tau$ the assumption of common trend at level 2 ($\psi_2 = \psi_1$) is now required in addition to common trend at level 1 ($\beta_2 = \beta_1$).

\subsection{General remarks}
\label{sec:generalremarks}
Whether the treatment acts at individual or cluster level determines the type of confounder to be accounted for unbiased estimation of the treatment effect. In fact, for an individual-level treatment (Figure \ref{graph:2a}) the confounder is the individual latent ability $A_{ij}$, whereas for a cluster-level treatment (Figure \ref{graph:2b})
the confounder is the cluster-level latent ability $\bar{A}_{j}$.

Therefore, the model equation of the conditioning approach should not include the cluster mean of the pre-test score if the treatment is at the individual level. In fact, as noted at the end of Section \ref{sec:individual-treat}, controlling for the cluster mean of the pre-test does not help reducing the bias since the cluster mean of the latent ability is not a confounder. However, it can introduce measurement error bias.  On the other hand, when the treatment is at the cluster level, the cluster mean of the latent ability is a confounder, thus the cluster mean of the pre-test must be included.

As regard the gain score approach, for unbiased estimation of the treatment effect the common trend assumption is needed only at level 1 if the treatment is at the individual level, whereas it is needed also at level 2 if the treatment is at the cluster level.

A strategy to caution against violations of the common trend assumption is fitting the gain score model with either the pre-test score $Y_{1ij}$ in place of $A_{ij}$ or the cluster mean of the pre-test score $\bar{Y}_{1j}$ in place of $\bar{A}_{j}$. In general, this strategy is not recommended because it may introduce a measurement error in the gain score model. However, when the treatment is at cluster level, introducing the cluster mean of the pre-test score $\bar{Y}_{1j}$ is advisable in case of clusters of large size since the measurement error is negligible. In fact, the reliability of $\bar{Y}_{1j}$ is an increasing function of the cluster size that rapidly approaches one (see equation (\ref{eq:Between_reliab}) of Appendix A).

The bias formulas for the single-level case (Section \ref{sec:method}) cannot be easily extended to the multilevel case since estimators are not in closed form. However, in a linear model the inclusion of random effects does not change the meaning of the regression coefficients since marginal and conditional coefficients coincide. Adding random effects just implies a modification of the estimation procedure which typically results in similar point estimates and larger standard errors \citep{snijders:2011}. Therefore, we expect that the bias formulas for the single-level case are good approximations for the multilevel case. Anyway, an investigation of this point needs a simulation study.

\section{Simulation study}
\label{sec:simulation}
In a multilevel setting the estimators do not have closed form expressions, thus a suitable simulation study is needed to evaluate the performance of the conditioning and gain score approaches illustrated in Section \ref{sec:multilevel}.
First, we illustrate the main characteristics of the scenarios adopted in the study and provide details about the data generation and the model fitting. Finally, we discuss the main results. 

The goal of the simulation study is to assess the bias of the estimator of the treatment effect under different conditions, checking whether the bias formulas derived for the single-level setting (Section \ref{sec:method}) are good approximations. 

As in Sections \ref{sec:individual-treat} and \ref{sec:cluster-treat}, we consider two scenarios: (\emph{i}) treatment at individual level (path diagram of Figure \ref{graph:2a}); (\emph{ii}) treatment at cluster level (path diagram of Figure \ref{graph:2b}). In both cases the clusters at the time of the post-test are formed before treatment assignment and they are the same at pre-test and post-test.

Both scenarios are analyzed under different conditions involving the measurement error and the common trend assumption. More in detail, under Scenario 1 and Scenario 2 we consider the assumptions defined by the parameter specification described in Table \ref{tab:sub}.

The design of the simulation study mimics data on student achievement as measured by the Italian National Institute for the Evaluation of Educational system (Invalsi; \cite{invalsi:2018,invalsi:2019}). More specifically, we were inspired by the problem of estimating the effect of being in a private versus a public school on the test scores measured at the last year of lower secondary school (8th grade), accounting for pre-test scores measured at the end of primary school (5th grade), see
\cite{asa:2019}.

\begin{table}[!ht]
\caption{Assumptions considered under Scenario 1 and Scenario 2}
\label{tab:sub}     
\footnotesize  
\begin{tabular}{llcccc}
\hline\noalign{\smallskip}
&&\multicolumn{2}{c}{Measurement error}&\multicolumn{2}{c}{Common trend} \\ \cline{3-6}
& Description& $\lambda_1=0$  & $\lambda_2=0$   & $\beta_1  = \beta_2$ & $\psi_1  = \psi_2$    \\ \hline
I& No meas. error;  common trend              &$\checkmark$     & $\checkmark$    &$\checkmark$   & $\checkmark$  \\
II& Meas. error on $Y_1$;  common trend       &     & $\checkmark$    &$\checkmark$   & $\checkmark$   \\
III & No meas. error;  common trend lev 2     &    $\checkmark$ & $\checkmark$    &   & $\checkmark$   \\
IV & No meas. error;  common trend lev 1     &    $\checkmark$ & $\checkmark$    & $\checkmark$   &   \\
V & Meas. error on $Y_1$;  common trend lev 2&    & $\checkmark$    &   & $\checkmark$   \\
VI &Meas. error on $Y_1$;  common trend  lev 1 &    & $\checkmark$    &   $\checkmark$&  \\  
VII & No meas. error;  no common trend         &   $\checkmark$  & $\checkmark$    & &  \\  
VIII & Same meas. error on $Y_1, Y_2$; common trend &  $(1)$  & $(1)$    & $\checkmark$&$\checkmark$  \\  
 \hline
\multicolumn{6}{l}{$(1)$ \footnotesize{$\lambda_1=\lambda_2 \neq 0$} }
\end{tabular}
\end{table}

\subsection{Data generation}
The simulation study is based on $K = 1000$ data sets having a balanced hierarchical structure: each data set is composed of $N = 10,000$ individuals uniformly distributed in  $J = 100$ groups, thus, $n_j = n = 100$ for all $j$.   

When the treatment is at cluster level the confounder is $\bar{A}_j$ and the reliability of $\bar{Y}_{1j}$ depends on $n$ (see equation (\ref{eq:Between_reliab}) of Appendix A). Thus, in Scenario 2 we performed simulations also with $n=4$.

The data generation process is based on the steps described in the following.
\begin{description}
\item[Step 1. Generation of individual ability.] 
For each subject,  the ability $A_{ij}$ is generated from a Normal distribution
\begin{equation*}
A_{ij}  \sim N(0, \sigma^2_A), \quad \sigma^2_A = 1.
\end{equation*}
\item[Step 2. Allocation of individuals in groups.] The 25\% of individuals is randomly  assigned to the groups; the remaining  75\% is assigned  on the basis of the individual ability: (\emph{i}) individuals are  ordered according to $A_{ij}$,  (\emph{ii}) then they are picked up in groups of $n \times 0.75$, i.e. 75 when $n=100$ and 1 when $n=4$. After the allocation of subjects in clusters, we compute the cluster means of abilities $\bar{A}_{j}$, defined in Section \ref{sec:individual-treat}.
\item[Step 3. Treatment assignment.] 
 The binary treatment variable  is generated at individual level (Scenario 1) according to a Bernoulli distribution
$$
Z_{ij} \sim {\rm Bernoulli}(\pi_{ij}), 
$$
with $\pi_{ij} = Pr(Z_{ij} = 1\mid A_{ij})$ denoting the probability of being treated for individual $i$ in group $j$. The probability $\pi_{ij}$ is computed using the logit model of equation (\ref{eq:logit}), setting $\alpha = 1$ and $\delta  = \log\frac{0.2}{0.8} = -1.38629$  to obtain about 20\% of treated individuals, similar to the Invalsi data structure for the individual level treatment given by enrolment in a public vs a private school. 

Similarly, when the treatment acts at cluster level (Scenario 2), the binary treatment variable is generated at group level according to 
$$
Z_{j} \sim {\rm Bernoulli}(\pi_{j}), 
$$
with $\pi_{j} = Pr(Z_{j} = 1\mid \bar{A}_{j})$ denoting the probability of being treated for  group $j$. The probability $\pi_j$ is computed along the same lines as  equation \eqref{eq:logit}, with the cluster-mean ability $\bar{A}_{j}$ instead of the individual ability $A_{ij}$. 
\item[Step 4. Generation of  pre-test and post-test scores.] 
We generate the  independent  error terms as follows, where $i = 1, \ldots, n_{j}; j = 1, \ldots, J$:
\begin{align*}
e_{ij} & \sim N(0, \sigma^2_e), \quad \sigma^2_e=1  \\
v_{ij} & \sim N(0, \sigma^2_{v}), \quad \sigma^2_{v}=1\\
(u_{1j},u_{2j})'& \sim BN(\mathbf{0}, \bm \Sigma) \quad \bm \Sigma=\left(%
\begin{array}{cc}
  \sigma^2_{u_1} & \sigma_{12}  \\
\sigma_{12}    & \sigma^2_{u_2} \\
\end{array}%
\right) 
\end{align*}
We fixed $\sigma^2_{u_1} =\sigma^2_{u_2}= 1$ and $\sigma_{12}=0.8$; note that when $\sigma_{12}=1$ the cluster-level errors $u_{1j}$ and $u_{2j}$ are perfectly correlated.

Both pre-test and post-test scores  depend on the individual ability, whose effect is different at within and between levels. 
The individual pre-test scores are generated using the random intercept model  \eqref{eq:gen1:P} under both scenarios,
while the post-test scores are generated using the random intercept model  \eqref{eq:gen1:Y} for Scenario 1 and model \eqref{eq:gen1:Y2} for Scenario 2. 

In order to mimic the structure of Invalsi data, the parameter values used in the simulations are set as follows: $\mu_1 = \mu_2 = 60$, $\tau = 2$, $\beta_1=16$. The values of  $\beta_2$, $\psi_1$, $\psi_2$, $\lambda_1$, and $\lambda_2$ depend on the specific configuration I-VIII of  Table \ref{tab:sub},
 as reported in Tables \ref{tab:res_tau} and \ref{tab:res_tau2}.

As concerns the value assigned to $\lambda_1$ and $\lambda_2$, we fix  $\lambda_1=\lambda_2 = 0$ under configurations I, III, IV,and VII (no measurement error) that imply perfect reliability; while we fix $\lambda_1 = 6$ under configurations II, V, VI and VIII (measurement error on the pre-test), leading to a reliability $\rho=0.88$, which is the value observed in the 5th-level Invalsi pre-test. For more details about the computation of  reliability in the multilevel setting see Appendix A. For configuration VIII we set  $\lambda_1=\lambda_2=6$.

The within-level parameter $\beta_2$  is set to $16$ (equal to $\beta_1$) in configurations I, II, 
IV, VI and VIII (common trend holds at level 1), while $\beta_2$ is set to 24 in configurations III, V and VII (common trend violated at level 1). 
We set the contextual parameter $\psi_1=8$ in configuration VII, where common trend assumption is violated at both levels; while $\psi_1=0$ in all other configurations.  Finally, the contextual parameter $\psi_2$ is set to zero in configurations I, II, III, V, VIII (common trend at level 2), while $\psi_2=8$ in configurations 
IV, VI, where the common trend at level 2 is violated, and $\psi_2=4$ in configuration VII (common trend violated at levels 1 and 2). 
It is worth to note that in configurations where $\psi_1=\psi_2$ we obtain the same results for $\psi_1=\psi_2=0$ (no contextual effect) and $\psi_1=\psi_2=8$ (contextual effects at the same extent on pre-test and post-test). This is the reason why in Tables \ref{tab:res_tau} and \ref{tab:res_tau2} we only report $\psi_1=\psi_2$.

\end{description}

\subsection{Main results}
The model parameters are estimated using random intercept linear models. The post-score $Y_{2}$ is regressed on the treatment variable $Z$ and the pre-test $Y_1$. In order to evaluate the role of the sample cluster means  $\bar Y_{1j}$, we considered two model specifications: (\emph{i}) without $\bar Y_{1j}$ and (\emph{ii}) with $\bar Y_{1j}$.

All models are fitted using the NLMIXED procedure of SAS \citep{SAS:2018}. 

We report the results of the simulation study in terms of  relative error (in \%) of the estimation of treatment parameter $\tau$
$$
\%err = \frac{\hat{\tau} - \tau}{\tau}*100,
$$
where $\hat{\tau}$ is the Monte Carlo (MC) mean of the estimated treatment effects over $1000$ simulations with \emph{true} $\tau = 2.0$.  

In what follows we first focus on results of Scenario 1 (Section \ref{sec:res_sc1}) and, then, we illustrate results of Scenario 2 (Section \ref{sec:res_sc2}).

\subsection{Results of Scenario 1}
\label{sec:res_sc1}

Results of the MC simulation study when the treatment acts at individual level (Scenario 1) are displayed in Table \ref{tab:res_tau}. For each configuration, we show the MC means of the estimated treatment effect together with the corresponding relative error, under the conditioning approach and the gain score approach.

Table \ref{tab:res_tau} reports only the results from models without $\bar Y_{1j}$. Indeed, the results are not affected by the inclusion of the sample cluster means $\bar Y_{1j}$, thus confirming what discussed in Section \ref{sec:generalremarks}.

\begin{table}[!ht] 
\caption{Scenario 1 (individual-level treatment): MC means of estimated treatment effect ($\hat \tau$) and corresponding relative error ($\%err$) of the estimated treatment effect,  under conditioning and gain score models; cluster size  $n_j=100$.}
\label{tab:res_tau}      
\footnotesize 
\begin{tabular}{lcccrrrrr}
    \hline
\multicolumn{4}{c}{Configuration}      &   \multicolumn{2}{c}{Conditional}     &   \multicolumn{2}{c}{Gain}   \\ \cline{5-8}
& & & &   $\hat \tau$    & \%err   &   $\hat \tau$ &   \%err   \\ \hline
& \multicolumn{1}{c}{Measurement}&  \multicolumn{2}{c}{Common trend}              &      &     &     &      \\  \cline{3-4}
& \multicolumn{1}{c}{error}             &  lev 1     & lev 2              &      &     &     &      \\ 
\hline
I   & $\lambda_1=\lambda_2=0$           & $\beta_1=\beta_2$&$\psi_1=\psi_2$&   $\mathbf{2.0}$   &   $\mathbf{0.0}$   &   $\mathbf{2.0}$  &  $\mathbf{0.0}$   \\
II  & $\mathbf{\lambda_1 = 6}$     & $\beta_1=\beta_2$& $\psi_1=\psi_2$ &   3.9   &  95.5   &   $\mathbf{2.0}$  &   $\mathbf{0.0}$   \\
III & $\lambda_1=\lambda_2=0$  &     $\mathbf{\beta_2 = 24}$       & $\psi_1=\psi_2$ &   $\mathbf{2.0}$   &   $\mathbf{0.0}$   &   9.0  & 348.1   \\
 IV & $\lambda_1=\lambda_2=0$  &      $\beta_1=\beta_2$    &   $\mathbf{\psi_1 = 0,\; \psi_2 = 8}$ &   {\bf 2.0}  &   {\bf 0.0}  &   {\bf 2.0}  &{\bf 0.0}   \\
V  & $\mathbf{\lambda_1 = 6}$     &     $\mathbf{\beta_2 = 24}$       & $\psi_1=\psi_2$ &   4.9   & 143.0   &   9.0  & 348.5   \\
VI & $\mathbf{\lambda_1 = 6}$     & $\beta_1=\beta_2$&      $\mathbf{\psi_1 = 0,\; \psi_2 = 8}$      &   3.9   &  95.5   &   $\mathbf{2.0}$  &   $\mathbf{0.0}$   \\
VII  & $\lambda_1=\lambda_2=0$          &      $\mathbf{\beta_2 = 24}$      &     $\psi_1 = 8,\; \psi_2 = 4$       &   $\mathbf{2.0}$   &   $\mathbf{0.0}$   &   9.0  & 348.5   \\
VIII & $\mathbf{\lambda_1=\lambda_2 =6}$& $\beta_1=\beta_2$& $\psi_1=\psi_2$ &   $\mathbf{2.0}$   &   $\mathbf{0.0}$   &   $\mathbf{2.0}$  &   $\mathbf{0.0}$    \\ 
\noalign{\smallskip}\hline
\end{tabular}
\end{table}

Looking at the results we can confirm in the multilevel context the validity of the results obtained by \cite{kim:steiner:2019} (configurations I and II). Indeed, the treatment effect is correctly estimated whatever the approach when the pre-test is measured without error (reliability equals 1) and the common trend assumption is fully satisfied both at level 1 and level 2 (configuration I). Moreover, under the common trend assumption, the gain score approach has to be preferred to the conditioning one in the presence of measurement error on the pre-test (configuration II), also when the reliability is quite high (0.88), as  usual  when validated tests are used (e.g., with Invalsi data). 

On the other hand, if the common trend assumption holds only at level 2, such as in configurations III and V, the treatment effect obtained with the gain score approach is dramatically  overestimated.  In addition, when the pre-test is affected by measurement error (configuration V), the treatment effect estimated under the conditioning approach is biased too,  even if at a smaller extent ($\%err = 143\%$ instead of $348.5\%$). 

Note that, if the common trend assumption is violated only at level 2 (configurations IV and VI), the treatment effect is correctly estimated under both the approaches when the measurement error is absent (configuration IV) and under the gain score approach  in the presence of measurement error (configuration VI). Moreover, when the common trend assumption does not hold at both levels (configuration VII) the estimated treatment effect shows the same bias as in configurations III and V, thus stressing the relevance of the common trend assumption at level 1.

Configuration VIII shows what happens when the measurement error is present both on the pre-test and the post-test. In particular, when the measurement error has the same impact on $Y_1$ and $Y_2$ and the common trend assumption holds, the  treatment effect is correctly estimated with both the approaches, as shown in Section \ref{sec:common_error}, equation \eqref{eq:collbias2} for the standard linear model (no multilevel). If the common trend is removed or the impact of the measurement error differs between pre-test and post-test, the unbiasedness of the $\tau$ estimator    under the conditioning approach is lost (results not shown here). 

We performed many other simulations to check the effect of other parameters on the treatment effect estimation, but findings were less interesting (results available upon request). For instance, as expected, increasing  the effect of the ability on the treatment assignment the relative errors increase dramatically in both approaches.  On the contrary, changing the values of other parameters did not lead to different results on the treatment effect estimation, such as considering a higher proportion of treated subjects ($30\%$ instead of $20\%$) 
or forming groups on the basis of individual ability in different proportions (instead of  $75\%$ of subjects allocated not at random and 25\% at random).

Table \ref{tab:summary}  summarizes our findings, giving hints  on the preferred approach under different assumptions, when the treatment is at individual level.

\begin{table}[!ht]
\caption{Approaches yielding unbiased estimation of treatment effect under different assumptions.}
\label{tab:summary}      
\footnotesize 
\begin{tabular}{clcc|}
                   &          & \multicolumn{2}{c}{\emph{Measurement error on pre-test} $Y_1$ ($\lambda_1 \ne 0$)}  \\ 
                   &          & \emph{Yes}          &\multicolumn{1}{c}{\emph{No}}\\ \cline{3-4}
\emph{Common trend}&\emph{No}  & \multicolumn{1}{|c|}{\emph{\textbf{none}}}                                     & Conditional  \\ \cline{3-4}
\emph{at level 1}  &           & \multicolumn{1}{|c|}{Gain}                                  & Gain          \\  
($\beta_1=\beta_2$)&\emph{Yes} & \multicolumn{1}{|c|}{Conditional if $\lambda_1=\lambda_2$ } & \multicolumn{1}{|c|}{Conditional}     \\
                   &           & \multicolumn{1}{|c|}{(common measurement error)}            &   \\   \cline{3-4}  
\end{tabular}
\end{table}
\subsection{Results of Scenario 2}
\label{sec:res_sc2}

Results of the MC simulation study when treatment acts at group level (Scenario 2) are displayed in  Table \ref{tab:res_tau2}. For each configuration, we show the MC means of the estimated treatment effect together with the corresponding relative error, under the conditioning approach and the gain score approach. We also distinguish between models without and with the cluster-mean pre-test $\bar{Y}_{1j}$ among the explanatory variables.

\begin{table}[!ht] 
\caption{Scenario 2 (group-level treatment): MC means of estimated treatment effect ($\hat \tau$) and corresponding relative error ($\%err$) of the estimated treatment effect,  under conditioning and gain score models; cluster size $n_j=100$ and $n_j=2$.}
\label{tab:res_tau2}      
\footnotesize 
\begin{tabular}{lcccrrrrrrrrr}
    \hline
\multicolumn{4}{c}{Configuration}      &   \multicolumn{4}{c}{Conditional}     &   \multicolumn{4}{c}{Gain}   \\ \hline
& \multicolumn{1}{c}{Measurement}&  \multicolumn{2}{c}{Common trend}        &      \multicolumn{2}{c}{Without $\bar{Y}_{1j}$} &  \multicolumn{2}{c}{With $\bar{Y}_{1j}$}   &     \multicolumn{2}{c}{Without $\bar{Y}_{1j}$} &  \multicolumn{2}{c}{With $\bar{Y}_{1j}$}      \\ \cline{3-12} 
& \multicolumn{1}{c}{error}             &  lev 1     & lev 2 &   $\hat \tau$    & \%err   &   $\hat \tau$ &   \%err &  $\hat \tau$    & \%err   &   $\hat \tau$ &   \%err  \\ \hline
\multicolumn{2}{l}{$n_j=100$}& &&   &      &     &     &   &   &   &   \\  
I   &  $\lambda_1=\lambda_2=0$          &$\beta_1=\beta_2$&$\psi_1=\psi_2$ &   $\mathbf{2.0}$   &   $\mathbf{0.0}$   &   $\mathbf{2.0}$  &  $\mathbf{0.0}$   &   $\mathbf{2.0}$   &   $\mathbf{0.0}$   &   $\mathbf{2.0}$  &  $\mathbf{0.0}$ \\
II  & $\lambda_1 = 6$     &$\beta_1=\beta_2$&$\psi_1=\psi_2$&   4.2   &  108.0   &   $\mathbf{2.0}$  &   $\mathbf{0.0}$ & $\mathbf{2.0}$  &   $\mathbf{0.0}$ & $\mathbf{2.0}$  &   $\mathbf{0.0}$   \\
III &  $\lambda_1=\lambda_2=0$&    $\beta_2=24$        &$\psi_1=\psi_2$&   $\mathbf{2.0}$   &   $\mathbf{0.0}$   &   $\mathbf{2.0}$  &   $\mathbf{0.0}$ & 6.2 & 209.5 &  $\mathbf{2.0}$  &   $\mathbf{0.0}$  \\
 IV &  $\lambda_1=\lambda_2=0$ &  $\beta_1=\beta_2$   & $\psi_1=0, \, \psi_2=8$ &   6.3  &   213.0  &   {\bf 2.0}  &{\bf 0.0}   & 6.3 & 213.0 & $\mathbf{2.0}$  &   $\mathbf{0.0}$\\
V  & $\mathbf{\lambda_1 = 6}$  &    $\beta_2=24$        &$\psi_1=\psi_2$&   5.0   & 148.0   & $\mathbf{2.0}$  &   $\mathbf{0.0}$ &   6.2  & 209.0 & $\mathbf{2.0}$  &   $\mathbf{0.0}$   \\
VI & $\mathbf{\lambda_1 = 6}$     &$\beta_1=\beta_2$&       $\psi_1=0, \, \psi_2=8$     &   8.0   &  300.0   &   $\mathbf{2.0}$  &   $\mathbf{0.0}$  &   6.1 & 203.5 & $\mathbf{2.0}$  &   $\mathbf{0.0}$  \\
VII  &  $\lambda_1=\lambda_2=0$       &   $\beta_2=24$         &     $\psi_1=8, \, \psi_2=4$       &   $-3.0$   &   $-247.5$   &  $\mathbf{2.0}$  &   $\mathbf{0.0}$  &   $4.0$   &   $101.0$   &  $\mathbf{2.0}$  &   $\mathbf{0.0}$  \\
VIII & $\mathbf{\lambda_1=\lambda_2 = 6}$&$\beta_1=\beta_2$&$\psi_1=\psi_2$&   $\mathbf{2.0}$   &   $\mathbf{0.0}$   &   $\mathbf{2.0}$  &   $\mathbf{0.0}$    &   $\mathbf{2.0}$   &   $\mathbf{0.0}$   &   $\mathbf{2.0}$  &   $\mathbf{0.0}$\\ 
\noalign{\smallskip}\hline
\multicolumn{2}{l}{$n_j=4$}& &&   &      &     &     &   &   &   &   \\  
I &  $\lambda_1=\lambda_2=0$  &$\beta_1=\beta_2$ & $\psi_1=\psi_2$ & $\mathbf{2.0}$  & $\mathbf{0.0}$   &   $\mathbf{2.0}$  &  $\mathbf{0.0}$   &   $\mathbf{2.0}$         & $\mathbf{0.0}$  & $\mathbf{2.0}$   &  $\mathbf{0.0}$ \\
II  & $\lambda_1 = 6$     &  $\beta_1=\beta_2$& $\psi_1=\psi_2$ &   3.3   &  65.1   &  $2.5$  &   $25.8$ &
                                                 $\mathbf{2.0}$  &   $\mathbf{0.0}$ & $2.5$  &   $25.8$   \\
III &  $\lambda_1=\lambda_2=0$ & $\beta_2=24$    & $\psi_1=\psi_2$ &   $\mathbf{2.0}$   &   $\mathbf{0.0}$   &   $\mathbf{2.0}$  &   $\mathbf{0.0}$ & 6.7 & 233.0 &  $\mathbf{2.0}$  &   $\mathbf{0.0}$  \\
 IV &  $\lambda_1=\lambda_2=0$ & $\beta_1=\beta_2$     & $\psi_1=0, \, \psi_2=8$ &   6.6  &   $282.2$  &   {\bf 2.0}  &{\bf 0.0}   & 6.6 & 230.7 & $\mathbf{2.0}$  &   $\mathbf{0.0}$\\
V  & $\lambda_1 = 6$    & $\beta_2=24$    &  $\psi_1=\psi_2$ &   3.9   & 96.9   & $2.8$  &   $40.0$ &   6.6  & 229.0 & $2.8$  &   $40.0$   \\
VI & $\lambda_1 = 6$     & $\beta_1=\beta_2$& $\psi_1=0, \, \psi_2=8$ &   7.91   &  295.3   &   2.8  &   40.0  &   6.6 & 230.5 & $2.8$  &   $40.0$  \\
VII  &  $\lambda_1=\lambda_2=0$         &      $\beta_2=24$     & $\psi_1=8, \, \psi_2=4$&   $0.22$   &   $-89.0$   &   {\bf 2.0}  &{\bf 0.0}   &   $4.30$   &   $114.9$   &   {\bf 2.0}  &{\bf 0.0}  \\
VIII & $\lambda_1=\lambda_2 = 6$ &   $\beta_1=\beta_2$ & $\psi_1=\psi_2$ &   $\mathbf{2.0}$   &   $\mathbf{0.0}$   &   $\mathbf{2.0}$  &   $\mathbf{0.0}$   &   $\mathbf{2.0}$   &   $\mathbf{0.0}$   &   $\mathbf{2.0}$  &   $\mathbf{0.0}$   \\ 
\noalign{\smallskip}\hline
\end{tabular}
\end{table}

Looking at the results where the cluster-mean pre-test score is not included among the model regressors, simulations suggest that the absence of measurement error is a necessary but not sufficient condition to correctly estimate the treatment effect under the conditioning approach. Indeed, the treatment effect is correctly estimated only when the measurement error on the pre-test is absent and common trend holds at level 2, as in configurations I and III. On the opposite,  the treatment effect is biased in  configuration IV, where the  measurement error is absent, but the  common trend is violated at level 2.  

The situation sharply changes when the cluster mean pre-test $\bar{Y}_{1j}$ is introduced as a regressor in the conditioning model and the cluster size is large ($n=100$). 
Indeed, in this case  the conditional model yields unbiased estimates also  for $\lambda_1 \neq 0$. This can be explained considering that when the treatment assignment depends on the  ability cluster mean $\bar{A}_j$ (Scenario 2), its measure $\bar{Y}_{1j}$ has a reliability that approximates one when the cluster size is large, whatever the value of $\lambda_1$, (see equation (\ref{eq:Between_reliab}) in Appendix A). Moreover, in configuration IV, where there is not measurement error but common trend at level 2 does not hold, the insertion of $\bar{Y}_{1j}$ is needed to avoid level 2 endogeneity \citep{grilli:2011}.
These considerations are confirmed when looking at simulation results for small cluster size. Indeed, when  $n=4$, adding the cluster mean pre-test $\bar{Y}_{1j}$ does not solve the measurement error problem (configurations II, V, VI) because the reliability of $\bar{Y}_{1j}$ does not approximate 1. On the other side, the inclusion of $\bar{Y}_{1j}$ is needed to solve the endogeneity problem of configurations IV and VII. 

Concerning the gain score approach when the cluster-mean pre-test score is not included among the model regressors,  simulation results show unbiased estimates of the treatment effect only when common trend is satisfied at both levels (configurations I and II), whereas  common trend holding only at level 2 (configurations III and V) or only at level 1 (configurations IV, VI, and VII) is not sufficient to guarantee  the unbiasedness if the model does not contain $\bar{Y}_{1j}$. However, the inclusion of $\bar{Y}_{1j}$ in the gain score model is dangerous when the cluster size is small ($n=4$) in presence of measurement error (configuration II), because $\bar{Y}_{1j}$ reintroduces measurement error in the model, as observed in Section \ref{sec:generalremarks}.

\section{Concluding remarks} 
\label{sec:conclusion}

In this paper we considered the estimation of a treatment effect on a test score in observational studies when a pre-test is available, comparing the alternative approaches known as conditioning and gain score. Our contribution mainly resides in analyzing the merits and drawbacks of the two approaches in a multilevel setting, which is relevant in education as students are typically nested into schools.

We reviewed the bias formulas of the estimators of the treatment effect under the two approaches and explicitly considered the case of a binary treatment. Then we considered a two-level data generating model, investigating peculiar issues such as the distinction between individual-level and cluster-level treatment, the role of the contextual effects of the confounder (e.g., peer effects), and multilevel versions of the reliability and the common trend assumption.

We exploited the structure of the model and the analytical formulas available in the single-level case to give hints on the bias of the estimators in the multilevel setting. Then we devised a simulation study to investigate the performance of the conditioning and gain score approaches under different scenarios.

For a treatment at individual level, our results confirm the findings of the single-level setting. The conditioning approach gives a biased estimator of the treatment effect whenever the pre-test is affected by measurement error, though the bias disappears if the pre-test and post-test scores are affected by a common source of error at the same extent. The gain score approach provides an unbiased estimator if the common trend assumption holds at the individual level, regardless of the assumption holding at cluster level. For an individual-level treatment, including the cluster mean of the pre-test score as a regressor is not recommended, as it introduces further measurement error without reducing the bias.

On the other hand, the findings for a treatment at cluster level are somewhat different. In fact, the cluster mean of the latent ability acts as a confounder, thus including the cluster mean of the pre-test score as a regressor helps reducing the bias. As in the previous case, the cluster mean of the pre-test score has the drawback of carrying some measurement error. However, the amount of measurement error crucially depends on the size of the clusters. Indeed, in settings with large clusters the reliability of the cluster mean of the pre-test score is close to one, thus its use is advantageous. In fact, the simulation results show that with clusters of size 100 using the cluster mean of the pre-test score gives unbiased estimators under both approaches, regardless of the assumptions on the measurement error and the common trend. As expected, the simulation results with clusters of size 4 show that when the cluster size is small the insertion of the cluster mean of the pre-test score does not completely eliminate the bias, though it is largely convenient in the conditioning approach. Differently, caution is required under the gain score approach: in fact, if the common trend assumption holds at both levels, the estimator of the treatment effect is unbiased without the cluster mean of the pre-test score, whereas including the cluster mean induces a bias due to measurement error. Overall, for a treatment at cluster level, unless there are strong arguments to support the common trend assumption, we recommend the conditioning approach with the cluster mean of the pre-test score.

In the present study we have considered a simple model without covariates, but the results can be extended to include covariates. Future work should investigate other settings, for example a treatment influencing also the pre-test and a scenario with clusters formed after the treatment. 

\bibliographystyle{apacite}
\def\baselinestretch{1.2}

\section*{Appendix A - Reliability in the multilevel setting}
\label{sec:Appendix}
In the classical test theory \citep{Lord:1968} the reliability of a test is defined as the ratio between the  variance of the ``true'' score and the variance of the ``observed'' score. In our case, from equation \eqref{eq:gen1:P2}, the true score is ${T}_{1ij} = \mu_1+\beta_1^W \tilde{A}_{ij} +\beta_1^B \bar{A}_j +u_{1j}$, with variance
\begin{eqnarray}
Var(T_{1ij}) & = & (\beta_1^W)^2 \sigma^2_W +(\beta_1^B)^2 \sigma^2_B+\sigma^2_{u_1},
\end{eqnarray}
where $\sigma^2_{u_1}=Var(u_{1j})$, while $\sigma^2_W$ and $\sigma^2_B$ are the within and between variances of the ability, respectively: 
$Var(A_{ij}) =\sigma^2_A = \sigma^2_W+\sigma^2_B$.
The observed score is the pre-test score $Y_{1ij}={T}_{1ij}+ \lambda_1 e_{ij}$, with variance
\begin{eqnarray}
Var(Y_{1ij})& = & Var(T_{1ij}) +  \lambda_1^2 \sigma^2_e,
\end{eqnarray}
where $\sigma^2_e=Var(e_{ij})$. 
Therefore, the \textit{overall reliability} of $Y_{1ij}$ is
\begin{equation}
\label{eq:totreliab}
\rho_{Y_1}  = \frac{Var(T_{1ij})}{Var(Y_{1ij})}= 1-\frac{\lambda_1^2\sigma^2_e}{Var(Y_{1ij})}.
\end{equation}
If the measurement error is null ($\lambda_1 = 0$), the reliability \eqref{eq:totreliab} equals one. 
 
In a multilevel setting, it is worth to define also the \textit{level 2 reliability}, namely the reliability of the cluster mean of the observed score $\bar{Y}_{1j}$ as a measure of the cluster mean of the true score $\bar{T}_{1j}=\mu_1 + \beta_1^B \bar{A}_{j}+u_j$ \citep{snijders:2011}:
\begin{equation}
\rho_{\bar{Y}_{1}}   =  \frac{(\beta_1^B)^2 \sigma^2_B+\sigma^2_{u_1}}{(\beta_1^B)^2 \sigma^2_B + \sigma^2_{u_1}+ \frac{(\lambda_1)^2}{n} \sigma^2_e},
\label{eq:Between_reliab}
\end{equation}
where $n$ is the cluster size, in our case the number of students per school. Equation (\ref{eq:Between_reliab}) shows that the reliability of the cluster mean $\bar{Y}_{1j}$ is an increasing function of $n$, approaching one regardless of the value of the parameter $\lambda_1$ regulating the measurement error. 

\end{document}